\documentclass[runningheads]{llncs}
\usepackage{graphicx}
\usepackage{multirow}
\usepackage{booktabs}
\usepackage{subfigure}

\usepackage[colorlinks]{hyperref}
\newcommand{\bheading}[1]{{\noindent{\textbf{#1}}}}

\begin{document}
\title{Rib Suppression in Digital Chest Tomosynthesis}
\author{Yihua Sun\inst{1} \and
Qingsong Yao\inst{3} \and
Yuanyuan Lyu\inst{4} \and 
Jianji Wang\inst{5} \and
Yi Xiao\inst{6} \and
Hongen Liao\inst{1} \and 
S. Kevin Zhou\inst{2}}
\authorrunning{Y. Sun et al.}

\institute{Department of Biomedical Engineering, School of Medicine, Tsinghua University, Beijing, China \and 
Medical Imaging, Robotics, and Analytic Computing Laboratory and Engineering (MIRACLE) Center, School of Biomedical Engineering \& Suzhou Institute for Advance Research, University of Science and Technology of China, Suzhou, China \email{s.kevin.zhou@gmail.com} \and
Key Lab of Intelligent Information Processing of Chinese Academy of Sciences (CAS),
Institute of Computing Technology, CAS, Beijing, China \and 
Z$^2$Sky Technologies Inc., Suzhou, China \and
Affiliated Hospital of Guizhou Medical University, Guizhou, China \and
Department of Radiology, Changzheng Hospital, Naval Medical University, Shanghai, China}
\maketitle

\begin{abstract}
Digital chest tomosynthesis (DCT) is a technique to produce sectional 3D images of a human chest for pulmonary disease screening, with 2D X-ray projections taken within an extremely limited range of angles. However, under the limited angle scenario, DCT contains strong artifacts caused by the presence of ribs, jamming the imaging quality of the lung area. Recently, great progress has been achieved for rib suppression in a single X-ray image, to reveal a clearer lung texture. We firstly extend the rib suppression problem to the 3D case at the software level. We propose a \textbf{T}omosynthesis \textbf{RI}b Su\textbf{P}pression and \textbf{L}ung \textbf{E}nhancement \textbf{Net}work (TRIPLE-Net) to model the 3D rib component and provide a rib-free DCT. TRIPLE-Net takes the advantages from both 2D and 3D domains, which model the ribs in DCT with the exact FBP procedure and 3D depth information, respectively. The experiments on simulated datasets and clinical data have shown the effectiveness of TRIPLE-Net to preserve lung details as well as improve the imaging quality of pulmonary diseases. Finally, an expert user study confirms our findings.

\end{abstract}

\section{Introduction} 
Digital chest tomosynthesis (DCT)  is a relatively novel imaging modality using limited angle tomography to provide the benefits of 3D imaging~\cite{molk2015digital,zhou2019handbook,zhou2021review}, which is reconstructed from a series of X-ray projections acquired within an extremely limited angle range~\cite{dobbins2009chest} using filtered back projection (FBP)~\cite{lauritsch1998theoretical}. DCT shares some tomographic benefits with computed tomography (CT) as an adjunct to a conventional chest radiography exam for diagnosing pulmonary disease, and carries promise in clinical decision making. Compared with conventional chest radiography, DCT greatly improves the performance in lung nodule detection~\cite{jung2012digital,machida2016whole}.
While compared with low-dose CT, DCT achieves competitive performance with lower cost and less radiation dosage~\cite{dobbins2009chest} on the diagnosis of lung cancer~\cite{terzi2013lung}. DCT has advantages in the detection of the early stages of COVID-19~\cite{miroshnychenko2020contrasts,yao2021label}, too.

However, the limited angle scenario of DCT leaves strong artifacts of ribs overlapped with lung textures, making it difficult for doctors to identify some lung disease contexts close to the rib artifacts.
Recently, great progress has been achieved for rib suppression in a single 2D X-ray radiograph. At the software level, DecGAN~\cite{li2019ribmiccai,li2020ribtmi} proposes a CycleGAN-based network that translates chest X-ray images to the simulated images. Furthermore, RSGAN~\cite{han2022gan} improves the performance by using a disentanglement network. For rib suppression in DCT at the device level, dual-energy chest tomosynthesis (DE-DCT)~\cite{dualtomo} can provide rib-free DCT by irradiating the tissues with two different energy levels of radiation, but exposes the patient to extra radiation doses. Therefore, rib suppression for DCT at the software level remains an important and unsolved problem.

\begin{figure}[t]
	\centering 	\scriptsize
	\includegraphics[width=0.65\textwidth]{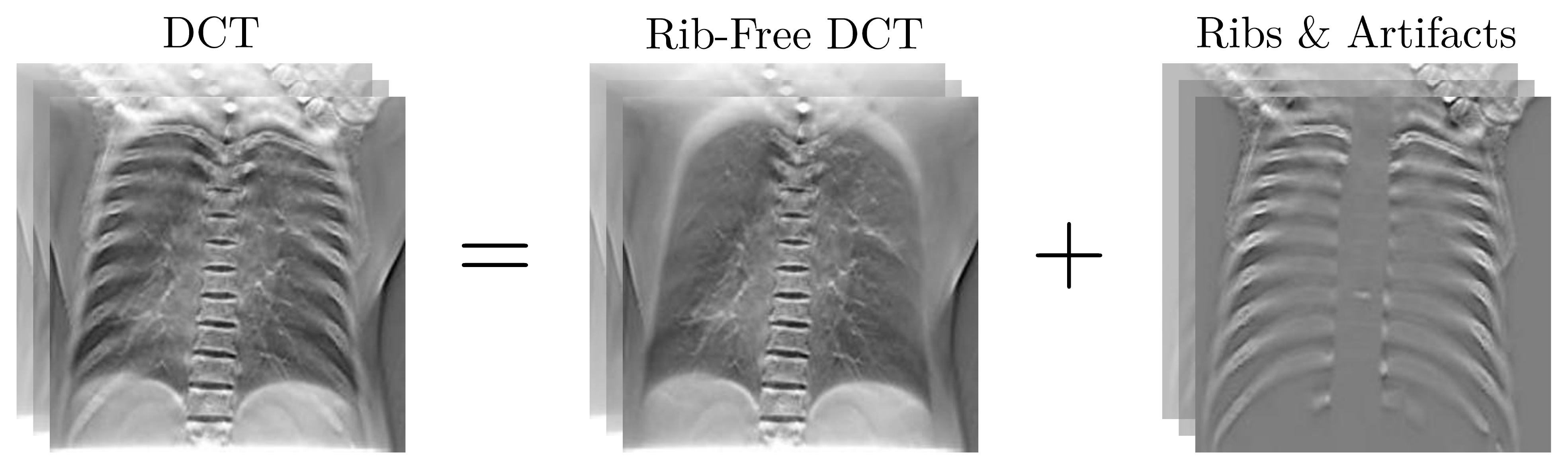}
	\caption{Visualizations of the DCT, rib-free DCT, and ribs with artifact.}
	\label{fig:dct_surpress_rib}
\end{figure}

We first visualize the difference between DCT and rib-free DCT in Fig.~\ref{fig:dct_surpress_rib}. The difference consists of the rib voxel itself and the artifacts generated by limited angle reconstruction using FBP, which should be removed simultaneously.
In this paper, we propose a \textbf{T}omosynthesis \textbf{RI}b Su\textbf{P}pression and \textbf{L}ung \textbf{E}nhancement \textbf{Net}work (TRIPLE-Net) to suppress the ribs and their artifacts in DCT without additive radiation dose. TRIPLE-Net leverages knowledge in both 2D X-ray projection images and 3D DCT reconstructed volumes with \textbf{three} convolutional neural networks:  projection-net, volume-net, and aggregation-net. 

The \underline{projection-net} suppresses rib component in the 2D projection, resulting in rib-free volume reconstructed by FBP, which is better at modeling artifacts based on the knowledge of FBP. However, decomposing rib components with high accuracy from the overlapped 2D projective textures might be difficult. Conversely, the \underline{volume-net} directly extracts the artifacts from lung tissues in 3D DCT volumes, which is expert in accurate and fine modeling. While our proposed \textbf{TRIPLE-Net} reaps the advantages of two sub-modules by merging their outputs with an \underline{aggregation-net}.

To the best of our knowledge, we are the first to suppress ribs along with their artifacts in DCT.
Extensive quantitative results, visualizations, and user studies validate that TRIPLE-Net can effectively suppress rib artifacts in DCT and keep the textures of lung tissues accurate and clear, which greatly outperforms competitive rib suppression methods in 2D X-ray radiograph.

\section{TRIPLE-Net}
\subsection{Problem Formulation}
The acquisition of DCT can be viewed as taking X-ray projection from multiple viewing angles within a limited range. The detector is stationary while the X-ray radiator moves vertically, which forms a cone-beam geometry. The collected X-ray projection logarithm can be expressed as,
\begin{equation}
\label{eq:proj}
I^\theta = -ln\int \eta(E)e^{-\mathcal{R}_\theta(f(E))}d E,
\end{equation}
where $f(E)$ is the 3D object's attenuation coefficient at energy level $E$, $\mathcal{R}_\theta$ is the 3D Radon transformation of DCT geometry at angle $\theta$, and $\eta$ is the energy distribution function of the X-ray spectrum.

Denote $I^\theta$ and $I_{rs}^\theta$ as the 2D X-ray projection at angle $\theta \in \Theta_\alpha$ and its corresponding rib-free projection, where $\Theta_\alpha=\{-\frac{\alpha}{2}, \cdots, \frac{\alpha}{2}\}$ is a set of angles in range of $\alpha$. The difference image $I_\Delta^\theta$ represents the rib component and artifacts: 
\begin{equation}
I_\Delta^\theta = I^\theta - I_{rs}^\theta.
\end{equation}

The 3D DCT volume $V$ and rib-free volume $V_{rs}$ reconstructed by a FBP operator $\mathcal{B}$ can be expressed as:
\begin{equation}
\label{eq:back proj}
V_\alpha = \sum_{\theta \in \Theta_\alpha}\mathcal{B}(I^\theta),\, V_{\alpha,rs} = \sum_{\theta \in \Theta_\alpha}\mathcal{B}(I_{rs}^\theta).
\end{equation}
Similarly, we denote the rib artifacts in 3D DCT as $V_{\alpha,\Delta}$. Given the linearity of FBP operator~\cite{lauritsch1998theoretical}, we have 
\begin{equation}
V_{\alpha,\Delta}= V_\alpha - V_{\alpha,rs} = \sum_{\theta \in \Theta_\alpha}\mathcal{B}(I^\theta) - \sum_{\theta \in \Theta_\alpha}\mathcal{B}(I_{rs}^\theta) = \sum_{\theta \in \Theta_\alpha}\mathcal{B}(I_\Delta^\theta).
\end{equation}

Accordingly, there are two approaches to suppress the rib artifacts in DCT: (\romannumeral1) remove rib component $I_\Delta^\theta$ from $I^\theta$ in 2D projection at each projection angle $\theta \in \Theta_\alpha$;  (\romannumeral2) or estimate $V_{\alpha,\Delta}$ directly from $V_\alpha$ in 3D.

\subsection{Method}
\begin{figure*}[t]
	\centering \scriptsize
	\includegraphics[width=0.9\textwidth]{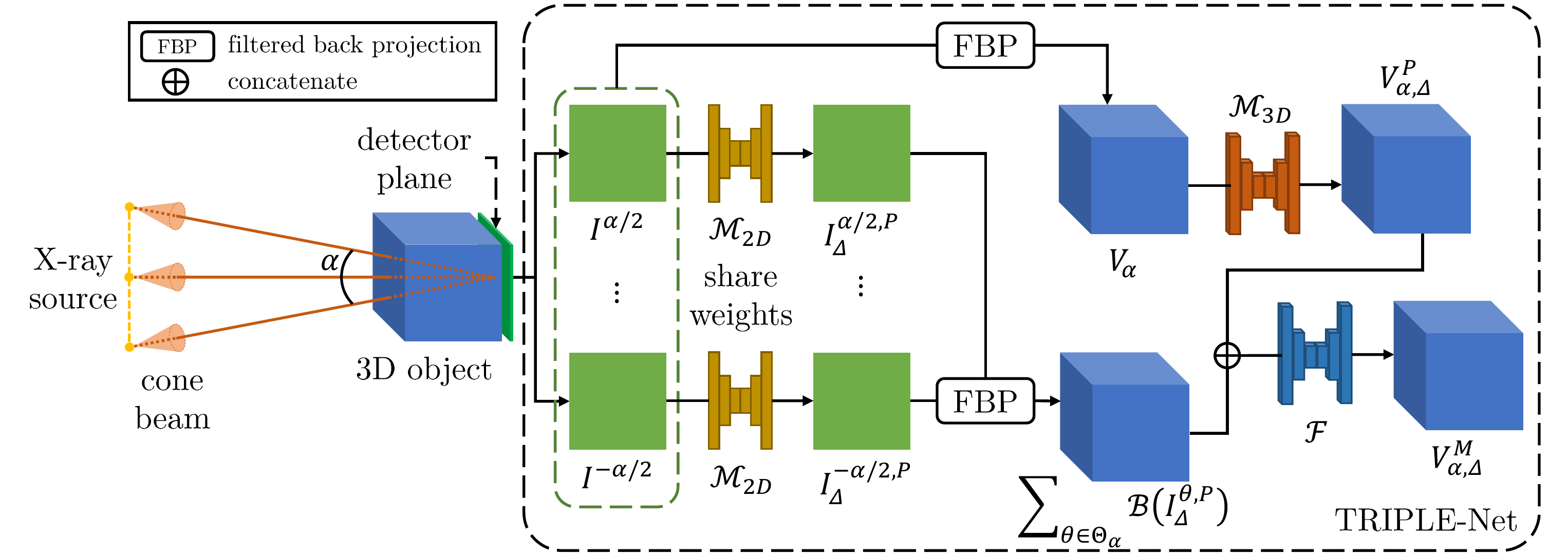}
	\caption{The overall framework of TRIPLE-Net. DCT has a cone-beam limited angle geometry. $\mathcal{M}_{2D}$, which shares weights across different views,  models rib components in the 2D domain before FBP. $\mathcal{M}_{3D}$  models rib components in the 3D domain after FBP. $\mathcal{F}$ takes the advantage of both worlds.}
	\label{fig:triplenet}
\end{figure*}

Figure~\ref{fig:triplenet} shows the framework of TRIPLE-Net, consisting of three convolutional neural networks: projection-net $\mathcal{M}_{2D}$, volume-net $\mathcal{M}_{3D}$, and aggregation-net $\mathcal{F}$. 

\bheading{2D rib component modeling.}
A 2D Residual U-Net (ResUNet)~\cite{kerfoot2018left} $\mathcal{M}_{2D}$ is utilized to extract $I_\Delta^{\theta,P} = \mathcal{M}_{2D}(I^\theta)$ from $I^\theta$. $\mathcal{M}_{2D}$ shares weights for each projection $I^\theta$ across all $\theta \in \Theta_\alpha$. The FBP operator $\mathcal{B}$ is employed to reconstruct 3D rib component in DCT with the 2D predictions $I_\Delta^{\theta,P}$ from all viewing angles. 

\bheading{3D rib component modeling.}
In the 2D projection domain, the rib component and other textures in lung tissues are overlapped. Besides, there is a lack of 3D consistency of the rib structures across different $\theta$. To more accurately model rib artifacts, a 3D ResUNet $\mathcal{M}_{3D}$ is employed to directly predict $V_{\alpha,\Delta}^P = \mathcal{M}_{3D}(V_\alpha)$, modeling rib artifacts in 3D reconstructed volume domain.

\bheading{Merging the results.}
$\mathcal{M}_{2D}$ and $\mathcal{M}_{3D}$ are able to suppress ribs in DCT based on the knowledge in 2D X-ray projections and 3D DCT volumes respectively. Here is the instinctive difference: $\mathcal{M}_{2D}$ is good at modeling artifacts, while $\mathcal{M}_{3D}$ is expert in modeling rib components with high accuracy.
Therefore, $\mathcal{F}$ is trained to learn the difference and leverages both the advantages of $\mathcal{M}_{2D}$ and $\mathcal{M}_{3D}$. A 3D ResUNet $\mathcal{F}$ is employed to merge the output of $\mathcal{M}_{2D}$ and $\mathcal{M}_{3D}$, producing the final prediction $V_{\alpha,\Delta}^M$ of ribs and their artifacts in DCT:
\begin{equation}
V_{\alpha,\Delta}^M = \mathcal{F}\left(\sum_{\theta \in \Theta_\alpha}{\mathcal{B}\left(I_\Delta^{\theta,P}\right)},\, V_{\alpha,\Delta}^P\right)
\end{equation}

\bheading{Loss functions.}
With the ground truth $I_\Delta^\theta$ and $V_{\alpha,\Delta}$, we perform supervised learning with the loss function defined as follows,
\begin{equation}
\mathcal{L}_{2D} = \lambda_{2D}\cdot||I_\Delta^{\theta,P}-I_\Delta^\theta||_1,\,\mathcal{L}_{3D} = \lambda_{3D}\cdot||V_{\alpha,\Delta}^P-V_{\alpha,\Delta}||_1.
\end{equation}
Since we have ground truth for both 2D and 3D domain, $\mathcal{M}_{2D}$ and $\mathcal{M}_{3D}$ are trained separately and the parameters are fixed before training $\mathcal{F}$. With the pretrained $\mathcal{M}_{2D}$ and $\mathcal{M}_{3D}$, the loss function for training $\mathcal{F}$ is defined as, 
\begin{equation}
\mathcal{L}_{M} = \lambda_{M}\cdot||V_{\alpha,\Delta}^M-V_{\alpha,\Delta}||_1
\end{equation}
We experimentally set the hyperparameters $\lambda_{2D}$, $\lambda_{3D}$ and $\lambda_{M}$ to 20, 50, and 50.

\section{Experiment}
\subsection{Experimental Setup}

\noindent\textbf{CT datasets.}
We use 4 online available CT datasets, LIDC-IDRI(``LI'')~\cite{lidc-idri,armato2011lung,clark2013cancer}, RibFrac (``RF'')~\cite{ribfrac2020}, MIDRC-RICORD-1a (``MR'')~\cite{midrc-ricord-1a,tsai2021rsna,clark2013cancer} and NSCLC Radiogenomics (``NR'')~\cite{nsclc,bakr2018radiogenomic,gevaert2012non,clark2013cancer}, to simulate DCT. As DCT requires a high resolution in the coronal plane, only CTs with spacing $\le 2.5mm$ in the longitudinal direction are selected. ``MR'' and ``NR'' have manually labeled masks for COVID-19 and lung cancer (mostly in nodule manifestation), which are utilized for testing the DCT image quality on lung disease. We use 90\% of ``LI'' and 90\% of ``RF'' for training, and 10\% of ``LI'', 10\% of ``RF'', ``MR'', and ``NR'' for testing. In total, there are 1353 training and 402 test data.

\noindent\textbf{Simulating DCT from CT.}
With CT, we have the 3D attenuation coefficient distribution of an object. We can simulate 2D X-ray projection images with our DCT geometry from CT using (\ref{eq:proj}) by deriving $f(E)$ from HU values. This procedure is known as the digitally reconstructed radiography (DRR) technique. We segment ribs in CT with a 3D U-Net~\cite{cciccek20163d} and inpaint rib mask with surrounding tissues, deriving the 3D attenuation coefficient distribution function denoted as $f_{rs}(E)$. Then we can obtain rib suppressed 2D X-ray projections $I_{rs}^\theta$ with DRR similarly. The volumes $V_\alpha$ and $V_{\alpha,rs}$ are accordingly reconstructed by FBP with the simulated projections as described in (\ref{eq:back proj}). In this paper, we do not focus on reconstruction algorithms for DCT, but on rib suppression in DCT; so we refer to $V_{\alpha,\Delta}$ and $V_{\alpha,rs}$ as ``ground truth'' for ribs on DCT and rib-free DCT.

\noindent\textbf{Implementation details.}
The 3D field of view (FOV) for DCT is set to $409.6mm\times300mm\times409.6mm$ with a shape of $256\times128\times256$, for the desired resolution of DCT in the anterior-posterior direction is lower. Before DRR,  the lung area of CT is placed at the center of FOV. The DRR and FBP procedures are implemented with ODL~\cite{adler2017operator}. The shape of $I^\theta$ and $I_{rs}^\theta$ are $256\times256$. We train and evaluate models separately for $\alpha=30^\circ$ and $\alpha=15^\circ$, with 59 and 29 projections taken equiangularly in the range of $\alpha$. $\mathcal{M}_{2D}$, $\mathcal{M}_{3D}$ and $\mathcal{F}$ are implemented with MONAI~\cite{monai} in PyTorch~\cite{pytorch} framework and trained with Adam~\cite{adam} optimizer with a learning rate of $1\times10^{-4}$ for 100 epochs.

\noindent\textbf{Performance metrics.} We segment the lung area $LA$ in CT with a lung mask~\cite{Hofmanninger2020}, and calculate $L_1$ and $L_2$ criteria within $LA$ of DCT ($L_1^{LA},\,L_2^{LA}$) and on the whole DCT volume ($L_1,\,L_2$) for evaluation over the whole test dataset. Besides, we use peak signal-to-noise ratio (PSNR) for evaluation within disease masks of ``MR'' and ``NR'' images pre-normalized to $[0,1]$.

\noindent\textbf{Clinical study.} We collect 4 clinical DCTs with $\alpha=15^\circ$ and 1 clinical DCT with $\alpha=30^\circ$ for evaluation, referred as the clinical dataset. We randomly select 30 cases from the test dataset and simulate DCTs with both $\alpha=15^\circ$ and $30^\circ$, resulting in 60 simulated DCTs. We invited 2 clinical doctors to give rankings from 1-5 (higher is better) for the DCTs by paying attention to the rib suppression performance and lung details. The DCT processed by different methods are randomly shuffled before presenting to the doctors. Doctor A is a proficient radiologist for chest imaging with over 20 years of reading experience. Doctor B is an orthopedist but with the knowledge of DCT. We use paired Wilcoxon signed-rank test deriving p-values to compare scores of other methods with TRIPLE-Net's.

\subsection{Comparison on simulated dataset}
\noindent\textbf{Quantitative metrics.} RSGAN~\cite{han2022gan} is a disentanglement method with generative adversarial networks designed for better performance on clinical chest X-ray images, which may have  deteriorated performances on the simulated dataset. Therefore, we train $\mathcal{M}_{2D}$ solely on the DRR domain to make a fair comparison of 2D and 3D methods on our simulated dataset.
Table~\ref{table:comp} shows the quantitative comparison of different methods, where TRIPLE-Net achieves the best performance. TRIPLE-Net has the best rib suppression ability on the whole DCT volume, and within the lung area which is the major concern of DCT. Moreover, TRIPLE-Net has better image quality where the pulmonary disease lies.
\begin{table}[t]
\centering \scriptsize 
    \caption{Quantitative comparison of different methods. TRIPLE-Net achieves the best performance in different regions and in terms of all metrics.}
	\begin{tabular}{c|l|c|c|c|c|c}
        \bottomrule \hline
		$\alpha$ & Method & $L_1(\times 10^{-2})$ & $L_2(\times 10^{-4})$ & $L_1^{LA}(\times 10^{-2})$ & $L_2^{LA}(\times 10^{-4})$ & $PSNR(dB)$\\
		\hline
		\multirow{4}{*}{$\;30^\circ\;$}
		&RSGAN & 0.853 & 3.272 & 1.891 & 6.997 & 25.09\\
		&$\mathcal{M}_{2D}$ & 0.753 & 2.406 & 1.793 & 6.344 & 25.07\\
		&$\mathcal{M}_{3D}$ & 0.662 & 1.590 & 1.033 & 2.280 & 33.37\\  
		&TRIPLE-Net & \textbf{0.471} & \textbf{0.948} & \textbf{0.869} & \textbf{1.658} & \textbf{33.56}\\
		\hline
		\multirow{4}{*}{$\;15^\circ\;$}
		&RSGAN & 1.164 & 5.957 & 2.641 & 13.45 & 25.11\\
		&$\mathcal{M}_{2D}$ & 1.013 & 4.323 & 2.446 & 11.51 & 25.46\\
		&$\mathcal{M}_{3D}$ & 0.873 & 2.756 & 1.640 & 5.356 & 30.52\\  
		&TRIPLE-Net & \textbf{0.687} & \textbf{2.058} & \textbf{1.397} & \textbf{4.200} & \textbf{30.85}\\
        \hline \toprule
	\end{tabular}
	\label{table:comp}
\end{table}

\begin{figure*}[h]
	\centering
	\scriptsize
	\begin{minipage}{0.8\textwidth}
		\begin{minipage}[t]{0.18\textwidth}
		    \vspace{2pt\centering Ground Truth$_{}$}
			\centering
			\includegraphics[width=1\textwidth]{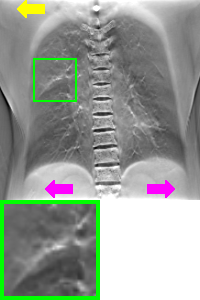}
		\end{minipage}
		\begin{minipage}[t]{0.18\textwidth}
		    \vspace{2pt\centering RSGAN$_{}$}
			\centering
			\includegraphics[width=1\textwidth]{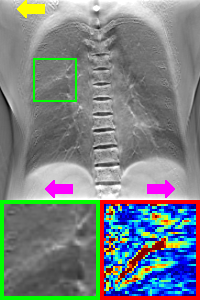}
		\end{minipage}
		\begin{minipage}[t]{0.18\textwidth}
			\vspace{2pt\centering $\mathcal{M}_{2D}$}
			\centering
			\includegraphics[width=1\textwidth]{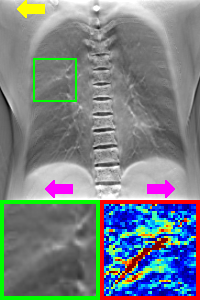}
		\end{minipage}
		\begin{minipage}[t]{0.18\textwidth}
			\vspace{2pt\centering $\mathcal{M}_{3D}$}
			\centering
			\includegraphics[width=1\textwidth]{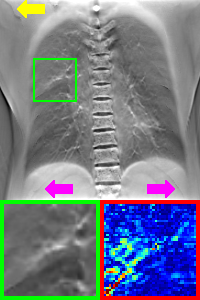}
		\end{minipage}
		\begin{minipage}[t]{0.18\textwidth}
		    \vspace{2pt\centering TRIPLE-Net$_{}$}
			\centering
			\includegraphics[width=1\textwidth]{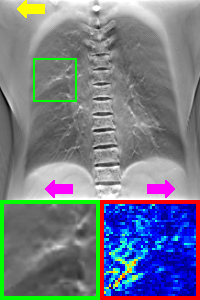}
		\end{minipage}
	\end{minipage}
	\\\vspace{2pt}
	\vspace{2pt\centering \leftline{\hspace{51pt}DCT\hspace{96pt}difference maps}}
	\begin{minipage}{0.8\textwidth}
		\begin{minipage}[t]{0.18\textwidth}
			\centering
			\includegraphics[width=1\textwidth]{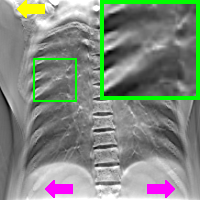}
		\end{minipage}
		\begin{minipage}[t]{0.18\textwidth}
			\centering
			\includegraphics[width=1\textwidth]{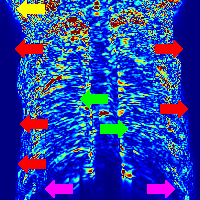}
		\end{minipage}
		\begin{minipage}[t]{0.18\textwidth}
			\centering
			\includegraphics[width=1\textwidth]{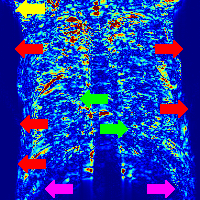}
		\end{minipage}
		\begin{minipage}[t]{0.18\textwidth}
			\centering
			\includegraphics[width=1\textwidth]{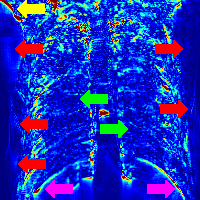}
		\end{minipage}
		\begin{minipage}[t]{0.18\textwidth}
			\centering
			\includegraphics[width=1\textwidth]{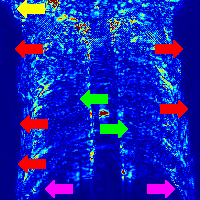}
		\end{minipage}
		\begin{minipage}[t]{0.04\textwidth}
    		\centering
    		\includegraphics[height=4.5\textwidth]{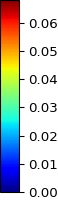}
    	\end{minipage}
	\end{minipage}
	\caption{Visualization of rib suppression results on DCT with $\alpha=30^\circ$. The methods with 3D information can better preserve lung details while removing rib components (green arrows/rectangles). The 2D methods only affect the ``rib-FBP'' area and no more, having better accuracy in ``non-rib'' and ``tricky-artifacts'' area outside lung area (red/yellow/pink arrows). TRIPLE-Net leverages the advantages of both 2D and 3D methods.}
	\label{fig:rib supp}
\end{figure*}
\noindent\textbf{Visualization on rib suppression performance.} Figure~\ref{fig:rib supp} shows the rib suppression performance of different methods on DCT with $\alpha=30^\circ$. The lung details are magnified in the green rectangles with a unified window level, whose corresponding difference maps are in the red rectangles. We can observe that with DCT rib suppression techniques, the lung details (green rectangles) intersecting with rib artifacts can be revealed more clearly.

Since 2D methods extract rib components in the 2D domain and model $V_{\alpha, \Delta}$ with the exact FBP procedure, they only affect the ``rib-FBP'' area and no more, as pointed by red arrows in Figure~\ref{fig:rib supp}. On the contrary, $\mathcal{M}_{3D}$ simply approximates the volumetric ground truth with a 3D neural network without the knowledge of the FBP mechanism, leaving error widely spread on the whole volume. Moreover, bone components are sometimes complex (bones around shoulders) or with lower contrast (ribs in tissue), as pointed by yellow and pink arrows in Figure~\ref{fig:rib supp}. So, 3D models have difficulty extracting features and modeling those tricky artifacts. 2D models are easier to model those tricky artifacts in DCT for they are caused by FBP and not that tricky in the 2D domain. With both 2D and 3D information, TRIPLE-Net effectively learns the modality difference by leveraging the advantages of both 2D and 3D domains and achieves the best visual quality with high accuracy within and beyond the lung area.

\begin{figure*}[ht]
	\centering
	\scriptsize
	\makebox[10pt][c]{$\quad$}
	\begin{minipage}{0.9\textwidth}
		\begin{minipage}[t]{0.13\textwidth}
		    \vspace{2pt\centering DCT}
		\end{minipage}
		\begin{minipage}[t]{0.13\textwidth}
		    \vspace{2pt\centering RSGAN}
		\end{minipage}
		\begin{minipage}[t]{0.13\textwidth}
		    \vspace{2pt\centering $\mathcal{M}_{2D}$}
		\end{minipage}
		\begin{minipage}[t]{0.13\textwidth}
		    \vspace{2pt\centering $\mathcal{M}_{3D}$}
		\end{minipage}
		\begin{minipage}[t]{0.13\textwidth}
		    \vspace{2pt\centering TRIPLE-Net}
		\end{minipage}
		\begin{minipage}[t]{0.13\textwidth}
		    \vspace{2pt\centering Ground Truth}
		\end{minipage}
		\hfill
		\begin{minipage}[t]{0.13\textwidth}
		    \vspace{2pt\centering CT}
    	\end{minipage}
	\end{minipage}
	\\\makebox[10pt][c]{(a)}
    \begin{minipage}{0.9\textwidth}
		\begin{minipage}[t]{0.13\textwidth}
			\centering
			\includegraphics[width=1\textwidth]{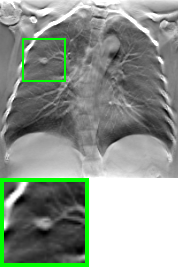}
		\end{minipage}
		\begin{minipage}[t]{0.13\textwidth}
			\centering
			\includegraphics[width=1\textwidth]{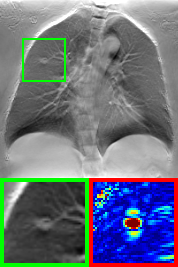}
		\end{minipage}
		\begin{minipage}[t]{0.13\textwidth}
			\centering
			\includegraphics[width=1\textwidth]{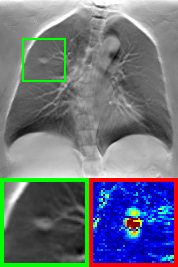}
		\end{minipage}
		\begin{minipage}[t]{0.13\textwidth}
			\centering
			\includegraphics[width=1\textwidth]{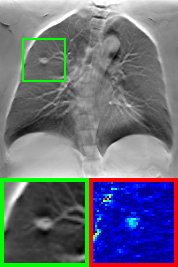}
		\end{minipage}
		\begin{minipage}[t]{0.13\textwidth}
			\centering
			\includegraphics[width=1\textwidth]{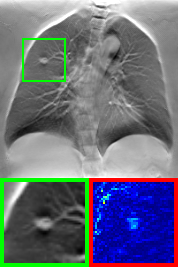}
		\end{minipage}
		\begin{minipage}[t]{0.13\textwidth}
			\centering
			\includegraphics[width=1\textwidth]{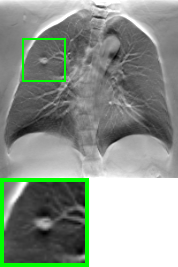}
		\end{minipage}
		\hfill
		\begin{minipage}[t]{0.13\textwidth}
    		\centering
    		\includegraphics[width=1\textwidth]{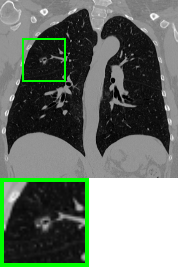}
    	\end{minipage}
	\end{minipage}
	\\\vspace{2pt}
	\makebox[10pt][c]{(b)}
	\begin{minipage}{0.9\textwidth}
	    \begin{minipage}[t]{0.13\textwidth}
			\centering
			\includegraphics[width=1\textwidth]{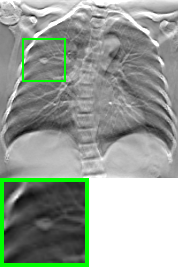}
		\end{minipage}
		\begin{minipage}[t]{0.13\textwidth}
			\centering
			\includegraphics[width=1\textwidth]{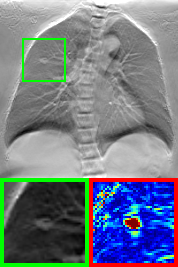}
		\end{minipage}
		\begin{minipage}[t]{0.13\textwidth}
			\centering
			\includegraphics[width=1\textwidth]{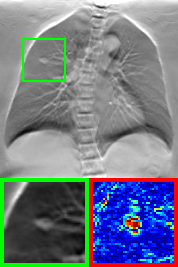}
		\end{minipage}
		\begin{minipage}[t]{0.13\textwidth}
			\centering
			\includegraphics[width=1\textwidth]{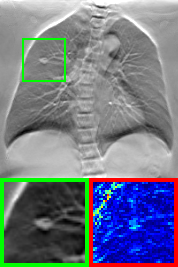}
		\end{minipage}
		\begin{minipage}[t]{0.13\textwidth}
			\centering
			\includegraphics[width=1\textwidth]{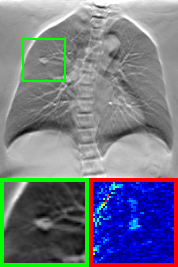}
		\end{minipage}
		\begin{minipage}[t]{0.13\textwidth}
			\centering
			\includegraphics[width=1\textwidth]{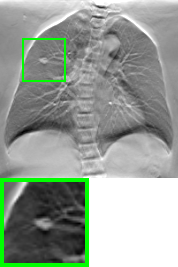}
		\end{minipage}
		\hfill
 		\begin{minipage}[b]{0.14\textwidth}
 		    \centering
			\vspace{2pt\scriptsize\centering(a)$\;\alpha=30^\circ$}
			\vspace{2pt\scriptsize\centering(b)$\;\alpha=15^\circ$}
			$\quad$\\
			$\quad$
		\end{minipage}
	\end{minipage}
	\caption{Lung nodules imaging performance of different methods. With DCT rib suppression techniques, lung details buried by rib artifacts are revealed more clearly. The 2D methods provide smoothed results while the 3D methods present the lung nodule with better contrast and sharper boundary. TRIPLE-Net has better accuracy within and around the lung nodule.}
	\label{fig:nodules}
\end{figure*}

\noindent\textbf{Visualization on disease imaging quality.} In Figure~\ref{fig:nodules}, images with lung nodules are visualized, where the affected area is magnified in the green rectangles with a unified window level for $\alpha=15^\circ$ and $30^\circ$, respectively. The corresponding original CT, from which the DCTs are simulated, is also presented. With DCT rib suppression techniques, lung nodules buried by rib artifacts can be identified more easily with clearer boundaries. In the result of 2D models, the lung nodule and lung texture are smoothed. The lung nodule is more distinguishable with better contrast in the result of 3D models. Besides, TRIPLE-Net has a more accurate intensity compared with $\mathcal{M}_{3D}$, within and around the diseased area.

\noindent\textbf{Reducing the DCT's acquisition angle range.} Because of a more limited range of angles, DCT with $\alpha=15^\circ$ contains stronger artifacts than $\alpha=30^\circ$. It is harder to identify lung nodules when $\alpha=15^\circ$ than $\alpha=30^\circ$ as shown in Figure~\ref{fig:nodules}.
But with TRIPLE-Net, visually the DCT with $\alpha=15^\circ$ has a comparative imaging quality as $\alpha=30^\circ$ for lung textures, especially lung nodules. This shows that TRIPLE-Net has the potential to further reduce the DCT acquired angle range. It is beneficial to scenarios that do not allow for a large source movement, \textit{e.g.}, intraoperative imaging. Furthermore, it also carries a promise in reducing a radioactive dose, which is harmful to patients.

\subsection{Clinical study}
\begin{figure*}[ht]
	\centering  \scriptsize	
    \makebox[30pt][c]{$\alpha=30^\circ$}
	\begin{minipage}{0.8\textwidth}
		\begin{minipage}[t]{0.19\textwidth}
		    \vspace{2pt\centering FBP$_{}$}
			\centering
			\includegraphics[width=1\textwidth]{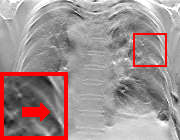}
		\end{minipage}
		\begin{minipage}[t]{0.19\textwidth}
		    \vspace{2pt\centering RSGAN$_{}$}
			\centering
			\includegraphics[width=1\textwidth]{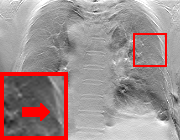}
		\end{minipage}
		\begin{minipage}[t]{0.19\textwidth}
		    \vspace{2pt\centering$\mathcal{M}_{2D}$}
			\centering
			\includegraphics[width=1\textwidth]{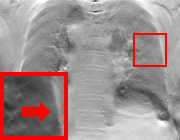}
		\end{minipage}
		\begin{minipage}[t]{0.19\textwidth}
		    \vspace{2pt\centering$\mathcal{M}_{3D}$}
			\centering
			\includegraphics[width=1\textwidth]{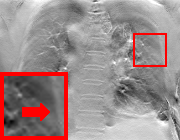}
		\end{minipage}
		\begin{minipage}[t]{0.19\textwidth}
		    \vspace{2pt\centering\mbox{TRIPLE-Net$_{}$}}
    		\centering
    		\includegraphics[width=1\textwidth]{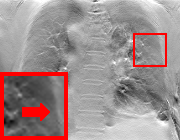}
    	\end{minipage}
	\end{minipage}
	\\\vspace{3pt}
	\makebox[30pt][c]{$\alpha=15^\circ$}
	\begin{minipage}{0.8\textwidth}
		\begin{minipage}[t]{0.19\textwidth}
			\centering
			\includegraphics[width=1\textwidth]{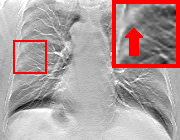}
		\end{minipage}
		\begin{minipage}[t]{0.19\textwidth}
			\centering
			\includegraphics[width=1\textwidth]{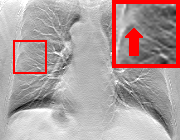}
		\end{minipage}
		\begin{minipage}[t]{0.19\textwidth}
			\centering
			\includegraphics[width=1\textwidth]{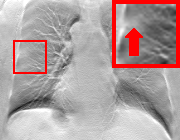}
		\end{minipage}
		\begin{minipage}[t]{0.19\textwidth}
			\centering
			\includegraphics[width=1\textwidth]{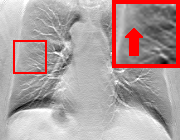}
		\end{minipage}
		\begin{minipage}[t]{0.19\textwidth}
    		\centering
    		\includegraphics[width=1\textwidth]{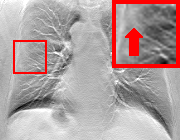}
    	\end{minipage}
	\end{minipage}
	\caption{Visualization of 2 different patients from clinical data, with $\alpha=15^\circ$ and $\alpha=30^\circ$. $\mathcal{M}_{2D}$ provided smoother lung textures. 3D models can better suppress rib artifacts in DCT compared with RSGAN, pointed by red arrows.}
	\label{fig:clinical}
\end{figure*}

\textbf{Visualization.} We directly run our trained model on clinical data from two patients and visualize the results in Fig.~\ref{fig:clinical}. To better generalize TRIPLE-Net to clinical data, RSGAN is substituted for the 2D sub-module $\mathcal{M}_{2D}$. The noticeable difference is contained in the red rectangle, being magnified and showed in a unified window level respectively for $\alpha=15^\circ$ and $\alpha=30^\circ$.
RSGAN preserves more lung detail than $\mathcal{M}_{2D}$, for its stronger generalization ability in the clinical data. However, without cross-view or 3D information, there is still some rib component leftover in the result of RSGAN (red arrows). Lung details in the red rectangle in the result of $\mathcal{M}_{2D}$ are smoothed. With 3D information, models can better remove rib components and preserve lung details on the whole image.

\noindent\textbf{Expert rating.} In Table~\ref{table:rating}, TRIPLE-Net has a score greater than other models with a statistical significance on the simulated dataset. On clinical dataset, models with 3D information have slightly better performance than other methods as in Table~\ref{table:rating}, which motivates us to further improve our method in future. 

\begin{table}[h]
\centering \scriptsize
    \caption{Ratings of clinical and simulated DCT images processed by different methods. For comparison of TRIPLE-Net with other methods, p-values are also summarized.}
	\begin{tabular}{l|cc|cc|cc|cc}
        \bottomrule \hline
		&\multicolumn{4}{c|}{Clinical dataset }     &\multicolumn{4}{c}{Simulated dataset} \\
		\hline
		\multirow{3}{*}{Method} &\multicolumn{2}{c|}{Doctor A} &\multicolumn{2}{c|}{Doctor B} &\multicolumn{2}{c|}{Doctor A} &\multicolumn{2}{c}{Doctor B}\\
		&Rating &p-value &Rating &p-value &Rating &p-value &Rating &p-value\\
		\hline
		FBP &1.00$\pm$0.00 &0.031 &1.00$\pm$0.00 &0.031 &1.00$\pm$0.00 &$<$0.001 &1.02$\pm$0.13 &$<$0.001 \\
		RSGAN &2.20$\pm$0.40 &0.031 &2.00$\pm$0.00 &0.031 &2.38$\pm$0.49 &$<$0.001 &2.03$\pm$0.26 &$<$0.001 \\
		$\mathcal{M}_{2D}$ &3.00$\pm$0.63 &0.031 &3.00$\pm$0.00 &0.031 &2.92$\pm$0.78 &$<$0.001 &3.02$\pm$0.34 &$<$0.001 \\
		$\mathcal{M}_{3D}$ &4.40$\pm$0.80 &0.500 &\textbf{4.60$\pm$0.49} &0.688 &4.20$\pm$0.81 &0.057 &4.18$\pm$0.53 &$<$0.001 \\
		\hline
		TRIPLE-Net &\textbf{4.40$\pm$0.49} &n.a. &4.40$\pm$0.49 &n.a.  &\textbf{4.50$\pm$0.65} &n.a. &\textbf{4.75$\pm$0.43} &n.a. \\
        \hline \toprule
	\end{tabular}
	\label{table:rating}
\end{table}
\section{Conclusion}
In this paper, we have proposed TRIPLE-Net to model rib artifacts in DCT caused by limited angle FBP, leveraging information in both 2D and 3D domains and reaps the benefits from both worlds. TRIPLE-Net can suppress rib artifacts in DCT to obtain a clearer lung texture and better visualization of pulmonary disease areas, which has the potential for better diagnosis of lung nodules and COVID-19 in clinics. In future, research could be furthered for higher resolution in rib-suppressed DCT and better performance in clinical data.

\bibliographystyle{splncs04}
\bibliography{reference}

\begin{thebibliography}{10}
\providecommand{\url}[1]{\texttt{#1}}
\providecommand{\urlprefix}{URL }
\providecommand{\doi}[1]{https://doi.org/#1}

\bibitem{monai}
Medical open network for artificial intelligence (monai).
  \url{https://monai.io/}, accessed 27 Feb 2022

\bibitem{pytorch}
Pytorch. \url{https://pytorch.org/}, accessed 27 Feb 2022

\bibitem{adler2017operator}
Adler, J., Kohr, H., Oktem, O.: Operator discretization library (odl). Software
  available from \url{https://github. com/odlgroup/odl}  (2017)

\bibitem{armato2011lung}
Armato~III, S.G., McLennan, G., Bidaut, L., McNitt-Gray, M.F., Meyer, C.R.,
  Reeves, A.P., Zhao, B., Aberle, D.R., Henschke, C.I., Hoffman, E.A., et~al.:
  The lung image database consortium (lidc) and image database resource
  initiative (idri): a completed reference database of lung nodules on ct
  scans. Medical physics  \textbf{38}(2),  915--931 (2011)

\bibitem{lidc-idri}
Armato~III, S.G., McLennan, G., Bidaut, L., McNitt-Gray, M.F., Meyer, C.R.,
  Reeves, A.P., Zhao, B., Aberle, D.R., Henschke, C.I., Hoffman, E.A., et~al.:
  Data from lidc-idri [data set]. The Cancer Imaging Archive  (2015).
  \doi{https://doi.org/10.7937/K9/TCIA.2015.LO9QL9SX}

\bibitem{bakr2018radiogenomic}
Bakr, S., Gevaert, O., Echegaray, S., Ayers, K., Zhou, M., Shafiq, M., Zheng,
  H., Benson, J.A., Zhang, W., Leung, A.N., et~al.: A radiogenomic dataset of
  non-small cell lung cancer. Scientific data  \textbf{5}(1), ~1--9 (2018)

\bibitem{nsclc}
Bakr, S., Gevaert, O., Echegaray, S., Ayers, K., Zhou, M., Shafiq, M., Zheng,
  H., Zhang, W., Leung, A.N., Kadoch, M., et~al.: Data for nsclc radiogenomics
  collection. The Cancer Imaging Archive  (2017).
  \doi{http://doi.org/10.7937/K9/TCIA.2017.7hs46erv}

\bibitem{cciccek20163d}
{\c{C}}i{\c{c}}ek, {\"O}., Abdulkadir, A., Lienkamp, S.S., Brox, T.,
  Ronneberger, O.: 3d u-net: learning dense volumetric segmentation from sparse
  annotation. In: International conference on medical image computing and
  computer-assisted intervention. pp. 424--432. Springer (2016)

\bibitem{clark2013cancer}
Clark, K., Vendt, B., Smith, K., Freymann, J., Kirby, J., Koppel, P., Moore,
  S., Phillips, S., Maffitt, D., Pringle, M., et~al.: The cancer imaging
  archive (tcia): maintaining and operating a public information repository.
  Journal of digital imaging  \textbf{26}(6),  1045--1057 (2013)

\bibitem{dobbins2009chest}
Dobbins~III, J.T., McAdams, H.P.: Chest tomosynthesis: technical principles and
  clinical update. European journal of radiology  \textbf{72}(2),  244--251
  (2009)

\bibitem{gevaert2012non}
Gevaert, O., Xu, J., Hoang, C.D., Leung, A.N., Xu, Y., Quon, A., Rubin, D.L.,
  Napel, S., Plevritis, S.K.: Non--small cell lung cancer: identifying
  prognostic imaging biomarkers by leveraging public gene expression microarray
  data—methods and preliminary results. Radiology  \textbf{264}(2),  387--396
  (2012)

\bibitem{han2022gan}
Han, L., Lyu, Y., Peng, C., Zhou, S.K.: Gan-based disentanglement learning for
  chest x-ray rib suppression. Medical Image Analysis p. 102369 (2022)

\bibitem{Hofmanninger2020}
Hofmanninger, J., Prayer, F., Pan, J., R{\"o}hrich, S., Prosch, H., Langs, G.:
  Automatic lung segmentation in routine imaging is primarily a data diversity
  problem, not a methodology problem. European Radiology Experimental
  \textbf{4}(1), ~50 (2020)

\bibitem{ribfrac2020}
Jin, L., Yang, J., Kuang, K., Ni, B., Gao, Y., Sun, Y., Gao, P., Ma, W., Tan,
  M., Kang, H., Chen, J., Li, M.: Deep-learning-assisted detection and
  segmentation of rib fractures from ct scans: Development and validation of
  fracnet. EBioMedicine  (2020)

\bibitem{jung2012digital}
Jung, H., Chung, M., Koo, J., Kim, H., Lee, K.: Digital tomosynthesis of the
  chest: utility for detection of lung metastasis in patients with colorectal
  cancer. Clinical radiology  \textbf{67}(3),  232--238 (2012)

\bibitem{kerfoot2018left}
Kerfoot, E., Clough, J., Oksuz, I., Lee, J., King, A.P., Schnabel, J.A.:
  Left-ventricle quantification using residual u-net. In: International
  Workshop on Statistical Atlases and Computational Models of the Heart. pp.
  371--380. Springer (2018)

\bibitem{adam}
Kingma, D.P., Ba, J.: Adam: {A} method for stochastic optimization. In: 3rd
  International Conference on Learning Representations (2015)

\bibitem{lauritsch1998theoretical}
Lauritsch, G., H{\"a}rer, W.H.: Theoretical framework for filtered back
  projection in tomosynthesis. In: Medical Imaging 1998: Image Processing.
  vol.~3338, pp. 1127--1137. International Society for Optics and Photonics
  (1998)

\bibitem{li2020ribtmi}
Li, H., Han, H., Li, Z., Wang, L., Wu, Z., Lu, J., Zhou, S.K.: High-resolution
  chest x-ray bone suppression using unpaired ct structural priors. IEEE
  transactions on medical imaging  \textbf{39}(10),  3053--3063 (2020)

\bibitem{li2019ribmiccai}
Li, Z., Li, H., Han, H., Shi, G., Wang, J., Zhou, S.K.: Encoding ct anatomy
  knowledge for unpaired chest x-ray image decomposition. In: International
  Conference on Medical Image Computing and Computer-Assisted Intervention. pp.
  275--283. Springer (2019)

\bibitem{machida2016whole}
Machida, H., Yuhara, T., Tamura, M., Ishikawa, T., Tate, E., Ueno, E., Nye, K.,
  Sabol, J.M.: Whole-body clinical applications of digital tomosynthesis.
  Radiographics  \textbf{36}(3),  735--750 (2016)

\bibitem{miroshnychenko2020contrasts}
Miroshnychenko, O., Miroshnychenko, S., Nevgasymyi, A., Khobta, Y.: Contrasts
  comparison of same cases of chest pathologies for radiography and
  tomosynthesis. In: 2020 International Symposium on Electronics and
  Telecommunications (ISETC). pp.~1--4. IEEE (2020)

\bibitem{molk2015digital}
Molk, N., Seeram, E.: Digital tomosynthesis of the chest: a literature review.
  Radiography  \textbf{21}(2),  197--202 (2015)

\bibitem{dualtomo}
Sone, S., Kasuga, T., Sakai, F., Hirano, H., Kubo, K., Morimoto, M., Takemura,
  K., Hosoba, M.: Chest imaging with dual-energy subtraction digital
  tomosynthesis. Acta Radiologica  \textbf{34}(4),  346--350 (1993)

\bibitem{terzi2013lung}
Terzi, A., Bertolaccini, L., Viti, A., Comello, L., Ghirardo, D., Priotto, R.,
  Grosso, M., Group, S.S., et~al.: Lung cancer detection with digital chest
  tomosynthesis: baseline results from the observational study sos. Journal of
  Thoracic Oncology  \textbf{8}(6),  685--692 (2013)

\bibitem{tsai2021rsna}
Tsai, E.B., Simpson, S., Lungren, M.P., Hershman, M., Roshkovan, L., Colak, E.,
  Erickson, B.J., Shih, G., Stein, A., Kalpathy-Cramer, J., et~al.: The rsna
  international covid-19 open radiology database (ricord). Radiology
  \textbf{299}(1),  E204--E213 (2021)

\bibitem{midrc-ricord-1a}
Tsai, E.B., Simpson, S., Lungren, M.P., Hershman, M., Roshkovan, L., Colak, E.,
  Erickson, B.J., Shih, G., Stein, A., Kalpathy-Cramer, J., et~al.: Data from
  the medical imaging data resource center - rsna international covid radiology
  database release 1a - chest ct covid+ (midrc-ricord-1a). Data from The Cancer
  Imaging Archive  (2022). \doi{https://doi.org/10.7937/VTW4-X588}

\bibitem{yao2021label}
Yao, Q., Xiao, L., Liu, P., Zhou, S.K.: Label-free segmentation of covid-19
  lesions in lung ct. IEEE transactions on medical imaging  \textbf{40}(10),
  2808--2819 (2021)

\bibitem{zhou2021review}
Zhou, S.K., Greenspan, H., Davatzikos, C., Duncan, J.S., Van~Ginneken, B.,
  Madabhushi, A., Prince, J.L., Rueckert, D., Summers, R.M.: A review of deep
  learning in medical imaging: Imaging traits, technology trends, case studies
  with progress highlights, and future promises. Proceedings of the IEEE
  (2021)

\bibitem{zhou2019handbook}
Zhou, S.K., Rueckert, D., Fichtinger, G.: Handbook of medical image computing
  and computer assisted intervention. Academic Press (2019)

\end{thebibliography}
\end{document}